# Social well-being of a sample of Iranian nurses: a descriptive-analytic study

**ABSTRACT**

Background: According to the World Health Organization's definition, health is a 'state of complete physical, mental, and social well-being and not merely the absence of disease and infirmity'. However, the social aspect of health, or social well-being, has not been attended to as equally as the other aspects. Social well-being is people's perceptions and experiences in social circumstances as well as the degree of successful responses to social challenges.

Objective: The aim of the study was to investigate the social well-being of a sample of Iranian nurses.

**Methods:** This was a descriptive-analytic study conducted in 2013. The study population consisted of all 1200 staff nurses working in all hospitals located in Ardabil, Iran. We invited a random sample of 281 practicing nurses to respond to the 33-item Keyes's Social Well-Being Questionnaire. Study data were analyzed by using SPSS, v. 19.0.

**Results:** Participants' mean score of social well-being was $105.45 \pm 15.87$. Social well-being was significantly related to participants' age, gender, work experience, satisfaction with working in hospital and with income, familiarity with nursing prior to entering it, official position, and type of employment. However, there was no significant relationship between nurses' social well-being and their marital status, their parents and spouses' educational status, as well as the type of hospital.

**Conclusion:** Nurses' social well-being deserves special attention. Effective well-being promotion strategies should be executed for promoting their social well-being particularly in areas of social integration and social acceptance. Moreover, nurses, particularly female nurses, need strong financial, emotional, informational, and social support for ensuring their social well-being.

**Keywords:** Social well-being, Health, Nurses, Iran

**INTRODUCTION:** Health is an abstract, complex, and multidimensional concept. Probably, the simplest definition of the concept is the presence of a sense of well-being and the absence of disease (Wills, 2011).The most comprehensive definition of health was provided by the World Health Organization (WHO)(Organization, 2006). The WHO defines health as a 'state of complete physical, mental, and social well-being and not merely the absence of disease and infirmity'. According to this definition, one of the aspects of health is social well-being (SWB). Keyes (1998) defined SWB as people's perceptions and experiences in social circumstances as well as the degree of successful responses to social challenges(C. L. M. Keyes, 1998). He proposed that SWB consists of five dimensions including,

- Social integration: individual's evaluation of the quality of relationships to the society and self;
- Social acceptance: individual's interpretation and acceptance of other people based on their character as well as the feelings of confidence and comfort in interacting with them;
- Social contribution: individual's evaluation of his/her own social value as well as belief in having something valuable to share with the society;
- Social actualization: individual's belief in the evolution of society and the possibility of progress and actualization through it;
- Social coherence: individual's perception of the quality, organization, and the soundness of the living world(Callaghan, 2008; Key-Roberts, 2009).

Based on these five dimensions, SWB is individuals' description of their perceptions and experiences of their well-being in the society as well as satisfaction with their own social structure and function(Key-Roberts, 2009; Law, Steinwender, & Leclair, 1998).

Despite the emphasis of the WHO on the social aspect of health as well as the vast amount of health-related advancements and interventions, SWB has not been regarded yet as equally as other aspects of health(Larson, 1993). In a review study, Callagan (2008) found that only a few studies have provided a theoretical basis for the concept of SWB(Callaghan, 2008). Keyes and Shapiro (2004) also noted that there are still considerable sociopolitical debates surrounding SWB(C. L. Keyes & Shapiro, 2004). Similarly, in our country, Iran, SWB has been recently taken into account and has been identified as a main priority of the Iranian healthcare system(Ministry of Health & Iran, 2009; Samii Mercede, 2011).

According to the WHO, health is among the basic human rights. Similarly, nurses, as humans who are at the forefronts of healthcare services, have the same right(Henderson, 2006, 2011) . Inattention toward nurses' health, particularly the social aspect of health, could seriously damage their SWB, prevent college applicants from entering nursing, negatively affect practicing nurses' job satisfaction, increase their turnover rate, cause them to leave the profession(Zarea, Negarandeh, Dehghan-Nayeri, & Rezaei-Adaryani, 2009) , and exacerbate the current crisis of heavy shortage of nursing staff (Farsi, Dehghan Nayeri, Negarandeh, & Broomand, 2010). These consequences, in turn, compromise the quality of nursing care and potentially undermine public health (Jannati, 2011; Joolaee S, 2006; Sultana, Riaz, Mehmood, & Khurshid, 2011) .

Jolaee et al. (2006) reported nursing poor social status and other social factors as the main reasons behind nurses' negative attitude towards the profession and their intention to leave it(Joolaee S, 2006). Consequently, despite the acute crisis of widespread unemployment and considerable employment opportunity in nursing, many young job applicants are reluctant to enter nursing, opt for leaving soon after entering it, or continue nursing practice without enough satisfaction and motivation(de Castro, Cabrera, Gee, Fujishiro, & Tagalog, 2009) . According to the World Federation of Occupational Therapists, job satisfaction and motivation—as the main components of SWB—are the fundamental rights of every employee. Accordingly, employers, organizations, and societies need to provide a supportive work environment that can improve employees' abilities and give them a sense of welfare and happiness(Hammell & Iwama, 2012) . These positive feelings enhance employees' health, well-being, job satisfaction, motivation, and productivity and create a balance between the goals of organization and the needs and demands of customers and workers(Avey, Luthans, Smith, & Palmer, 2010) .

Despite the paramount importance of SWB, there are limited data on Iranian nurses' SWB. We conducted this study to span this gap. The aim of the study was to investigate SWB of a sample of Iranian nurses.

## METHODS

Design

This was a descriptive-analytic study conducted in 2013.

Sample and setting

The study population consisted of all 1200 staff nurses working in all hospitals located in Ardabil, Iran. These hospitals included four teaching hospitals, a hospital affiliated to the Iranian Social Security Organization, and a private hospital. The sample size was calculated by using the following formula, $n = \frac{Nt^2 \times (pq)}{Nd^2 + t^2(pq)}$ (Cochran). Accordingly, with a confidence interval of 0.95, a $d$ of 0.05, and a $p$ of 0.4, the sample size was determined to be 282. We allocated a quota to each hospital proportionate to the number of nursing staffs working in it. Accordingly, we employed the random sampling method for recruiting the participants from different working shifts.

Instrument

The study data were collected by using a demographic questionnaire and the Keyes's Social Well-being Questionnaire(C. L. M. Keyes, 1998) . The demographic questionnaire consisted of ten questions regarding participants' demographic data, two questions regarding their satisfaction with working in hospital and their income, and one question about familiarity with nursing prior to entering it. The latter three questions were scored on a three-point likert scale. The points were 'Little', 'Moderate', and 'Very much'.

The Keyes's Social Well-Being Questionnaire (KSWBQ) is a five-dimension self-administered questionnaire comprising 33 items. The dimensions of KSWBQ are social integration (seven items), social acceptance (seven items), social actualization (seven items), social contribution (six items), and social coherence (six items). KSWBQ items are scored on a five-point likert scale on which 1 stands for 'Completely disagree' and 5 stands for 'Completely agree'. Accordingly, the possible range of the total score of KSWBQ is 33–165. Higher scores of the questionnaire reflect better SWB. Given the unequal numbers of items in the dimensions of KSWBQ, we calculated the means, rather than the sums, of each dimension to maintain the 1–5 metric of the item responses. Accordingly, meaningful comparison of scores across the dimensions was possible. Cicognani et al. (2008) reported a Coronbach's alpha of 0.88 for the questionnaire(Cicognani et al., 2008). In the current study, two nursing lecturers translated the original English version of KSWBQ into Persian. Then, we invited three nursing lecturers to assess the face and the content validity of the translated version. Finally, we asked 30 nurses to complete the questionnaire for the purpose of reliability assessment. The Coronbach's alpha was 0.87.

Data analysis
We employed the Statistical Package for Social Sciences (SPSS, v. 19.0) for data analysis. Initially, we examined the distribution of the study variables by using the Kolmogrov-Smirnov test. The results of this test revealed that all the variables had a normal distribution (P value > 0.05). Accordingly, we used the independent-samples t, the one-way analysis of variance (one-way ANOVA), the repeated measures ANOVA, and the Pearson's product moment correlation tests for analyzing the study data. We set the level of significance at below 0.05.

Ethical considerations
The institutional Review Board and the Ethics Committee of Tehran University of Medical Sciences, Tehran, Iran, approved the study. Prior to the study, we informed the participants about the aim of the study and ensured them that their information will be reported anonymously. Moreover, we guaranteed the confidentiality of their personal information. Returning the completely filled questionnaire was considered as participants' consent for participation.

**FINDINGS**
Totally, 281 staff nurses participated in the study. The mean and the standard deviation of our participants' age and work experience were 33.10±5.92 and 9.50±6.05, respectively Table 1 shows the study participants' demographic characteristics.
The mean of the participants' total score of KSWBQ was 105.45±15.87. The means and the standard deviations of the KSWBQ dimensions are shown in Table 2. The results of the repeated measures ANOVA test revealed that there was a significant difference among the mean item scores of KSWBQ dimensions. Accordingly, we used the Bonferoni's correction of p value for pairwise comparisons which revealed that the mean item score of the social contribution dimension was significantly higher than the other dimensions. Moreover, the mean item scores of the social integration and the social acceptance dimensions were significantly lower than the other dimensions (P value < 0.05; Table 2).
The results of the Pearson's product moment correlation test revealed that nurses' SWB was significantly and positively correlated with their age(r=0.187, P value = 0.002) and work experience (r=0.186, P value = 0.002). Moreover, the results of the independent-samples t test showed that the mean score of male nurses' SWB was significantly higher than female nurses (P value = 0.004; Table 1).
The results of the one-way ANOVA test and the Scheffe post-hoc test illustrated that the mean SWB score of nurses who had complete satisfaction with their income was significantly higher than both nurses with moderate satisfaction (P value = 0.001) and nurses with complete dissatisfaction (P value = 0.002). However, the mean SWB score of nurses with moderate satisfaction and nurses with complete dissatisfaction did not differ significantly (P value = 0.668). The results of the same test also revealed that the mean SWB score of nurses who were completely satisfied with working in hospital was significantly higher than nurses who were moderately satisfied (P value = 0.008) and nurses who were completely dissatisfied with hospital work (P value < 0.001). Additionally, the latter two groups also differed significantly from each other (P value = 0.013; Table 1).

The results of the one-way ANOVA and the Scheffe post-hoc tests also demonstrated that the mean SWB score of nurses who were completely familiar with nursing at the time of entering it was significantly higher than the nurses who were completely unfamiliar (P value = 0.004). However, the difference between the mean SWB scores of nurses with complete familiarity and nurses with moderate familiarity as well as the difference between the mean SWB scores of nurses with moderate familiarity and nurses with complete unfamiliarity were not statistically significant (P value = 0.167 and 0.070, respectively; Table 1).

The results of the one-way ANOVA test revealed that there was a significant difference among the mean scores of nurses who held different official positions (P value < 0.000). Accordingly, the results of the Scheffe post-hoc test showed that this difference was between the mean SWB scores of supervisors and staff nurses as well as head-nurses and staff nurses (P value = 0.000 and P value = 0.048, respectively). However, the difference between the mean SWB scores of supervisors and head-nurses not statistically significant (P value = 0.266Table 1).

The results of the one-way ANOVA and the Scheffe post-hoc tests demonstrated that the mean SWB score of nurses who secured formal lifelong employment was significantly higher than nurses with other types of employment (P value < 0.05; Table 1). Finally, the results of the one-way ANOVA test revealed that there was no significant relationship between nurses' mean SWB score and their marital status (P value = 0.211), the educational status of their fathers (P value = 0.523), mothers (P value = 0.206), and spouses (P value = 0.554), as well as the type of hospital (P value = 0.202).

**DISCUSSION**

The aim of the study was to investigate SWB of a sample of Iranian nurses working in six hospital settings located in Ardabil, Iran. The mean SWB score of the participating nurses was 105.45±15.87. Key-Roberts (2009) found that students' mean SWB score was 160.64±23.67. This huge difference between the mean SWB score of the two studies is due to the fact that Key-Roberts (2009) used a seven-point scale for scoring the KSWBQ items. If we recalculate the score reported by Key-Roberts (2009) by using a five-point scale, the score will be 114.74(Key-Roberts, 2009). The new score is also slightly higher than our participating nurses' mean SWB score. This difference can be attributed to the difference between the difficulties and the conditions of student life and nurses' working life.

We also found that male nurses' mean SWB score was significantly higher than female nurses. Key-Roberts (2009) also found that compared with female students, male students had a higher SWB mean score both before and after the study(Key-Roberts, 2009). Garrosa et al. (2010) reported a significant correlation between nurses' burnout and gender(Garrosa, Rainho, Moreno-Jiménez, & Monteiro, 2010). In a meta-analysis study, Purvanova and Muros (2010) also found that compared with men, women are more at risk for burnout-related health problems(Purvanova & Muros, 2010). On the other hand, it has been shown that burnout has a significant effect on different aspects of health (Gilbar, 1998).Accordingly, female nurses' lower SWB score is probably due to the more powerful effects of burnout on women's health(Emslie, Hunt, & Macintyre, 1999).

The study finding also revealed that nurses' mean SWB score was significantly correlated with their age and work experience. Arafa et al. (2003) and Van Lente et al. (2012) also found that age was a significant negative predictor of mental health(Arafa, Nazel, Ibrahim, & Attia, 2003; Van Lente et al., 2012). Khaghanizadeh et al. (2006) reported that mental health problems are more common in the first decade of nurses' working life(Khaghanizadeh, SIRATI, & Abdi, 2006). They reported a significant correlation between nurses' mental health status and their work experience. The significant correlation of SWB with age and work experience is probably due to the fact that older people usually have more working and real-world experiences and better cognitive, coping, and social skills and hence, are able to protect their health more effectively.

We also found a significant relationship between SWB and satisfaction with income. This is in line with the findings of Key-Roberts (2009)(Key-Roberts, 2009). Ferrer et al (2005) also highlighted the importance of income to people's welfare and happiness. They found that happiness and well-being were significantly correlated with income and perceived income equality (Ferrer-i-Carbonell, 2005). Other studies also reported that income level significantly affected psychosocial well-being (Ettner, 1996; Faragher, Cass, & Cooper, 2005; Shapiro & Keyes, 2008; Van Lente et al., 2012) .

The study findings revealed that nurses' SWB was significantly related to their satisfaction with working in hospital. Adults spend much of their time for working. Accordingly, work environment is an important aspect of people's life that significantly affects their job satisfaction and welfare(Faragher et al., 2005; Happell, Martin, & Pinikahana, 2003) . Harter et al. (2003) noted that satisfaction with work accounts for about 25% of the total variance of overall satisfaction with life(Harter, Schmidt, & Keyes, 2003). Schulz and Northridge (2004) also highlighted the impacts of environment, environmental stress, and social inequality on health and satisfaction(Schulz & Northridge, 2004). Given the world serious nursing staff shortage as well as the effect of nursing staffs' job satisfaction on patient satisfaction , work environment assumes an increasing importance(Zarea et al., 2009). Vahey et al. (2004) reported that environmental factors such as good staffing ratio, effective intra-professional relationships, and adequate support from managers significantly decrease nurses' burnout and increase patient satisfaction(Vahey, Aiken, Sloane, Clarke, & Vargas, 2004).

The study findings also demonstrated a significant relationship between nurses' SWB and their familiarity with nursing at the time of entering it. Generally, informed career choice is associated with more positive feelings, healthier interpersonal relationships, greater work-related happiness, pleasure, and productivity, and stronger loyalty towards costumers and organization(Harter et al., 2003) .

There was also a significant relationship between nurses' SWB and their official position and the type of employment. We found that supervisors and nurses who secured formal lifelong employment had higher SWB score. These findings imply that nurses who have greater job security have better SWB. Other studies also reported that official position and job security positively contribute to psychosocial aspects of employees' health(Kuhnert, Sims, & Lahey, 1989; Marmot & Wilkinson, 2005). Conversely, employees who are more worried about their social support and working condition are more at risk for developing health problems (Vahtera, Kivimäki, Pentti, & Theorell, 2000).

The study findings showed that our participants' SWB was not significantly related to the type of hospital. This finding contradicts the findings of a study conducted by Arafa et al. (2003) (Arafa et al., 2003). This contradiction can be attributed to the difference in the settings and cultural backgrounds of the two studies. Finally, we found no significant relationship between our participating nurses' SWB and their marital status. Arafa et al. (2003) and Shapiro and Keyes (2008) also reported the same finding(Arafa et al., 2003; Shapiro & Keyes, 2008). However, Sharbatriyan (2013) found a significant correlation between SWB and the marital status of a sample of Iranian students(sharbatriyan, 2013). This contradiction is probably due to the difference in the study populations of the two studies.

## CONCLUSION

The study findings indicate that nurses' mean SWB score is much lower than the maximum possible score (i.e. 165), implying that their SWB deserves special attention. Consequently, effective well-being promotion strategies should be developed and implemented for improving nurses' SWB particularly in areas of social integration and social acceptance. Moreover, given the significant correlation of SWB with income, work environment, prior knowledge of nursing before entering it, and perceived job security, providing nurses—particularly female nurses—with strong financial, emotional, informational, and social support may help improve their SWB.

### Limitations of the study

In this study, we had a tight budget and limited amount of time. Accordingly, we conducted the study only in one city. Therefore, the study findings might have limited generalizability.

### Acknowledgement

This study was funded by the Nursing and Midwifery Research Center of Tehran Medical University, Tehran, Iran. We would like to sincerely thank the study funder as well as the nurses who graciously agreed to participate in the study.